\documentclass[fleqn]{2017SCGE}
\allowdisplaybreaks[4]

\usepackage{amssymb}%
\usepackage{amsmath}
\usepackage{bm}

\begin{document}
\ensubject{subject}
\ArticleType{Article}
\SpecialTopic{SPECIAL TOPIC: }
\Year{2017}
\Month{January}
\Vol{60}
\No{1}
\DOI{10.1007/s11433-018-9213-7}
\ArtNo{000000}
\ReceiveDate{February 5, 2018}
\AcceptDate{March 28, 2018}

\title{Single particles in a reflection-asymmetric potential}{Single particles in a reflection-asymmetric potential}

\author[1]{WANG Yuanyuan}{}
\author[2]{REN Zhengxue}{zhxr1992@pku.edu.cn}

\AuthorMark{Wang Y Y }

\AuthorCitation{Wang Y Y, Ren Z X }

\address[1]{School of Physics and Nuclear Energy Engineering and International Research Center for Nuclei and Particles in the Cosmos, Beihang University, Beijing 100191, China}
\address[2]{State Key Laboratory of Nuclear Physics and Technology, School of Physics, Peking University, Beijing 100871, China}


\abstract{ Single particles moving in a reflection-asymmetric potential are investigated by solving the Schr\"{o}dinger equation of
the reflection-asymmetric Nilsson Hamiltonian with the imaginary time method in 3D lattice space and the harmonic oscillator basis expansion method.
In the 3D lattice calculation, the $\bm{l}^2$ divergence problem is avoided by introducing a damping function,
and the$\langle \bm{l^2}\rangle_N$ term in the non-spherical case is calculated by introducing an equivalent $N$-independent operator.
The efficiency of these numerical techniques is demonstrated by solving the spherical Nilsson Hamiltonian in 3D lattice space.
The evolution of the single-particle levels in a reflection-asymmetric potential is obtained and discussed by the above two numerical methods,
and their consistency is shown in the obtained single-particle energies with the differences smaller than 10$^{-4}~[\hbar\omega_0]$.}

\keywords{single particles, reflection-asymmetric potential, imaginary time method, harmonic oscillator basis expansion method}

\PACS{21.10.Pc, 21.60.Jz, 27.90.+b}

\maketitle

\begin{multicols}{2}

\section{Introduction}\label{sec_introduction}

Nuclear shape provides an intuitive understanding of spatial density distributions of atomic nuclei~\cite{bohr1975nuclear,ring2004nuclear},
and manifests itself in various exotic nuclear phenomena, such as halo phenomena in spherical~\cite{PhysRevLett.55.2676,PhysRevLett.77.3963,
PhysRevLett.80.460} and deformed nuclei~\cite{MENG2006470,PhysRevC.82.011301,meng2015halos}, superdeformation rotational bands~\cite{PhysRevLett.57.811},
tidal waves~\cite{frauendorf2011tidal,PhysRevC.93.044309}, chiral rotation~\cite{FRAUENDORF1997131,RevModPhys.73.463,meng2010open}, wobbling motion~\cite{bohr1975nuclear,PhysRevLett.86.5866} and nuclear fissions~\cite{zhou2016ps}.

The nuclei were assumed to be spherical in the 1950s~\cite{mayer1949}.
Based on the observations of the large quadrupole moments in some nuclei,
Rainwater pointed out that these nuclei might have deformed shapes~\cite{PhysRev.79.432}.
Then, Bohr and Mottelson proposed the nuclear collective model, which successfully explained the observed structure
\Authorfootnote
of the rotational bands~\cite{bohr1952quadrupole}.
In a more microscopic way, Nilsson extended the spherical shell model to a deformed case by introducing the degree of freedom for the quadrupole deformation~\cite{nilsson1955see}.
Since its introduction, the Nilsson model has achieved a great success in accounting for most of the observed features of single-particle levels in deformed nuclei,
and provided a proper microscopic basis to understand the properties of the nuclear structure.

With the first observation of negative-parity states in the even-even radium isotopes by Berkeley group in the 1950s~\cite{PhysRev.100.1543},
the possibility that nuclei might have a reflection-asymmetric shape arose~\cite{strutinsky1956zamechaniya,lee1957}.
In order to investigate these nuclei, one needs to calculate the single-particle energies in an average reflection-asymmetric potential.
Dutt and Mukherjee developed the reflection-asymmetric Nilsson Hamiltonian and investigated the single-particle motion in such potential~\cite{dutt1959}.
Based on the reflection-asymmetric Nilsson Hamiltonian,
many investigations have been performed~\cite{rustgi1963octupole,vogel1967nuclear,Moller1969223,LEANDER197593}.
The density functional theory~(DFT) provides an average potential in a self-consistent way and has achieved a great success in describing not only the single-particle motions~\cite{meng1999, chen2003cpl, zhou2003prl, Liang20151, meng2016relativistic, PhysRevLett.117.062502} but also the nuclear collective excitation modes~\cite{Afanasjev19991, Vretenar2005101, nikvsic2009beyond, li2009, zhao2011prl, zhao2016ps, meng2013progress, yao2014prc, yao2015prc, zhao2015prl, ZHOU2016227, PhysRevC.96.054324}.
A series of works on the reflection-asymmetric shaped nuclei have been done based on various DFTs for ground ~\cite{geng2007, wangnan2009, guo2010microscopic, huanghai2010, wei2010octupole, huangh2013, zhangwei2016} and excited states~\cite{bender2003rmf, lu2012potential, zhaojie2015prc, zhaojie2016prc, zhou2016ps}.
The DFTs are based on various effective nucleon-nucleon interactions.
The key problem to be solved in DFT is the single-particle motion in a mean-field potential with certain deformations.

In order to solve the equation for a single particle in a mean-field potential,
the harmonic oscillator (H.O.) basis expansion method has been widely used and achieved a great success~\cite{meng2016relativistic, meng2013progress, Liang20151}.
However, for nuclear phenomena with large spatial distribution, such as halo phenomena or nuclear fissions~\cite{MENG2006470},
a large basis space is needed to get a converged result.
In a more efficient way, one can solve the equations of the motion in a coordinate space.

The imaginary time method (ITM)~\cite{davies1980application,hagino2010iterative} is a powerful approach for solving the equations of motion in a coordinate space,
and it has been successfully applied to solve the Schr\"{o}dinger and Dirac equations~\cite{maruhn2014tdhf, zhangy2009cpc, zhangy2009cpl, zhangy2010ijmpe, hagino2010iterative, tanimura20153d, ren2017solving}.

In this paper, the ITM is used to solve the Schr\"{o}dinger equation of the reflection-asymmetric Nilsson Hamiltonian in 3D lattice space,
and the results obtained by the H.O. basis expansion method are compared with those from 3D lattice calculation.

The paper is organized as follows, the reflection-asymmetric Nilsson Hamiltonian,
together with the H.O. basis expansion method and the ITM in 3D lattice space will be briefly introduced in Sec.~\ref{sec_Theo}.
In Sec.~\ref{sec_num}, the reflection-asymmetric Nilsson Hamiltonian, and the numerical details
for the H.O. basis method and 3D lattice calculation are presented.
Several problems including the $\bm{l}^2$ divergence problem and the calculation of $\langle \bm{l^2}\rangle_N$ term are overcome by the introduced numerical techniques in Sec.~\ref{sec3Dchall},
and the efficiency of these treatments is also discussed.
Section~\ref{sec_sp} is devoted to discuss the single particles in a reflection-asymmetric potential.
Summary and perspectives are given in Sec.~\ref{sec_summary}.

\section{Theoretical framework}\label{sec_Theo}
\subsection{Reflection-asymmetric Nilsson Hamiltonian}

In present work, the single particles in a reflection-asymmetric potential are investigated by calculating the eigen energies
and eigen wave functions of the reflection-asymmetric Nilsson Hamiltonian~\cite{dutt1959},
\begin{align}
  \hat{h}= -\frac{\hbar^2}{2m}\Delta + V(r,\theta)+C\bm{l}\cdot \bm{s} + D[\bm{l}^2-\langle \bm{l}^2\rangle_N],\label{H}
\end{align}
with the kinetic energy term $-\hbar^2\Delta/2m$, the reflection-asymmetric potential $V(r,\theta)$, the spin-orbit term $C\bm{l}\cdot\bm{s}$,
the term $D[\bm{l}^2-\langle\bm{l}^2 \rangle_N]$, the orbit angular momentum $\bm{l}$, spin $\bm{s}$, and Nilsson parameters $C$ and $D$.

The reflection-asymmetric potential $V(r,\theta)$ has the form
\begin{align}
  V(r,\theta)
  & = \frac{1}{2}m\omega_0^2r^2
      \Big[
       1+a_1Y_{10}(\theta,\phi)+a_2Y_{20}(\theta,\phi)\notag\\
  &\quad\quad\quad\quad\quad\quad
        +a_3Y_{30}(\theta,\phi)
      \Big],\label{Vnilsson}
\end{align}
with parameters $a_1$, $a_2$ and $a_3$ describing the diople, quadrupole and octupole deformations, respectively. Assuming that the nuclear density is constant and distributes within an equipotential surface,
one can determine the relations among the parameters $a_1$, $a_2$ and $a_3$ by restricting the center of mass of nuclei coincided with the origin of the coordinate system and the volume conservation,
\begin{align}
  a_1 \sim \frac{9\sqrt{3}}{2\sqrt{35\pi}}a_2a_3,~~\omega_0^2 \sim \left[ 1+\frac{5}{16\pi}(a_2^2+a_3^2) \right]\mathring{\omega}^2_0,
\end{align}
where the higher order terms of $a_1, a_2$ and $a_3$ are neglected,
and $\mathring{\omega}_0$ corresponds to the frequency of the potential in the spherical case.
The parameters $a_2$ and $a_3$ can be related with the commonly used quadrupole and octupole deformation parameters $\beta_2$ and $\beta_3$,
\begin{align}
  a_2=-2\beta_2,~~ a_3=-2\beta_3.
\end{align}

The spin-orbit term  $C\bm{l}\cdot \bm{s}$ is essential to reproduce the right magic numbers~\cite{mayer1949}.
The $D[\bm{l}^2-\langle \bm{l}^2\rangle_N]$ term has the effect of interpolating between the oscillator and the square well and, thus, reproduces effectively the Woods-Saxon radial shape~\cite{nilssonbook}.
Here, $\langle \bm{l}^2\rangle_N=N(N+3)/2$ is the expectation value of $\bm{l}^2$ averaged over one spherical major shell with quantum number $N$.
In general, the constants $C$ and $D$ are given in the form of the Nilsson parameters $\kappa$ and $\mu$,
\begin{align}
  C=-2\hbar\mathring{\omega}_0\kappa,~~ D=-\hbar\mathring{\omega}_0\kappa\mu.
\end{align}
The values of $\kappa$ and $\mu$ depend on the mass number of the investigated nuclei, and one choice is ~\cite{nilsson1969nuclear},
\begin{align}
&
  \left.\begin{aligned}
    &\kappa_p=0.0766-0.0779\frac{A}{1000}\\
    &\mu_p=0.493+0.649\frac{A}{1000}
        \end{aligned}
  \right\}
\text{for ~protons ~and,}\\
&
  \left.\begin{aligned}
   &\kappa_n=0.0641-0.0026\frac{A}{1000}\\
   &\mu_n=0.624-1.234\frac{A}{1000}
        \end{aligned}
  \right\}
 \text{for ~neutrons.}
\end{align}

\subsection{The numerical methods}

In present work, two numerical methods are employed to obtain the eigen energies of the reflection-asymmertic Nilsson Hamiltonian.
One can diagonalize the Hamiltonian (\ref{H}) in a set of the spherical harmonic oscillator bases~$|Nlj\Omega\rangle$,
where $N$, $l$, $j$ and $\Omega$ represent the total number of the oscillator quanta, the orbit angular momentum, the total angular momentum and its $z-$component, respectively.

The ITM is an iterative method for solving the equations of the motion with arbitrary shapes.
The basic idea of the ITM is to replace time with an imaginary one, and the evolution of the wave function reads~\cite{davies1980application},
\begin{align}
  e^{-i\hat{h}t}|\psi_0\rangle \xrightarrow{t \rightarrow -i\tau} e^{-\hat{h}\tau}|\psi_0\rangle,
\end{align}
where $|\psi_0\rangle$ is an initial wave function and $\hat{h}$ is the Hamiltonian.

With the eigenstates $\{\phi_k\}$ of the Hamiltonian $\hat{h}$ corresponding to the eigenvalues $\{\varepsilon_k\}$,
the evolution of the wave function $|\psi(\tau)\rangle=e^{-\hat{h}\tau}|\psi_0\rangle$ can be written as
\begin{align}
  |\psi(\tau)\rangle = e^{-\hat{h}\tau}|\psi_0\rangle
                     = \sum_k e^{-\varepsilon_k\tau}|\phi_k\rangle \langle \phi_k|\psi_0\rangle,
\end{align}
where $\varepsilon_1\leq \varepsilon_2\leq\cdots$.
For $\tau \rightarrow\infty$, $|\psi(\tau)\rangle$ approaches the ground state wave function of $\hat{h}$ as long as $\langle \phi_1|\psi_0\rangle \neq 0$.

In the calculation, the imaginary time $\tau$ is discrete with the interval $\Delta \tau$, i.e., $\tau = N\Delta\tau$.
The wave function at $\tau=(n+1)\Delta\tau$ is obtained from the wave function at $\tau=n\Delta\tau$ by expanding the expotential evolution operator $e^{-\Delta\tau\hat{h}}$ to the linear order of $\Delta\tau$,
\begin{align}\label{Eq_pre_imaginary}
  |\psi^{(n+1)}\rangle = (1-\Delta\tau\hat{h})|\psi^{(n)}\rangle.
\end{align}
Since this evolution is not unitary, the wave function should be normalized at every step.

However, the convergence feature of Eq.~\eqref{Eq_pre_imaginary} is not so satisfactory as discussed in detail in Ref.~\cite{reinhard1982comparative}.
This is essentially because all energy components are propagated with the same step size $\Delta\tau$,
so that the actual evolution of each component is proportional to its energy.
This is clearly seen by expanding $|\psi^{(n)}\rangle$ in Eq.~\eqref{Eq_pre_imaginary} with the eigenstates $\{\phi_k\}$,
\begin{align}
   |\psi^{(n+1)}\rangle
= & (1-\Delta\tau\hat{h})|\psi^{(n)}\rangle\notag\\
= & \sum_k(1-\Delta\tau\varepsilon_k)|\phi_k\rangle\langle{\phi_k}|\psi^{(n)}\rangle.\label{phiiter}
\end{align}
Considering that the high energy components are usually dominated by the kinetic energy,
a simple iteration scheme Eq.\eqref{Eq_pre_imaginary} is improved by a kinetic-energy damping~\cite{reinhard1982comparative, BLUM1992comparison},
\begin{align}
   |\psi^{(n+1)}\rangle
= \left\{1-\frac{\delta}{\hat{T}+E_0}\left(\hat{h}-\langle\psi^{(n)}|\hat{h}|\psi^{(n)}\rangle\right)\right\}|\psi^{(n)}\rangle,
   \label{ekdamp}
\end{align}
where $\hat{T}=(\hbar^2/2m)\hat{\bm{p}}^2$ is the operator of kinetic energy, $\delta$ and $E_0$ are numerical parameters.
To get a stable and fast convergence, $E_0$ should be chosen typically of the same order of the depth of the potential, and $\delta$ is chosen from 0.1 to 0.8.
Of cause, a larger $\delta$ value leads to faster convergence of the evolution.
However, this makes the evolution meets the pathological conditions more easily.

To find excited states, one can start with a set of initial wave functions and orthonormalize them during the evolution procedure by the Gram-Schmidt method.
This method has been successfully employed in the 3D coordinate space calculations for nonrelativistic systems~\cite{maruhn2014tdhf}.


\section{Numerical details}\label{sec_num}

In the present work, the Nilsson parameters adopted are $\kappa=0.0589167$ and $\mu=0.640323$, which correspond to the proton potential with $A=227$.
For simplicity,  the natural units $\hbar=\omega_0=m=1$ are used.

For the calculation with the H.O. basis, a set of the spherical harmonic oscillator bases with $N_{\rm shell} = 20$ major shells are chosen in the diagonalization.
By increasing $N_{\rm shell}$ from 20 to 24, in the $(\beta_2,\beta_3)=(0.3,0.1)$ case, the single-particle energies below $7~[\hbar\omega_0]$ change less than $10^{-5}~[\hbar\omega_0]$.
This suggests that $N_{\rm shell} = 20$ is accurate enough.

For the 3D lattice calculation, the step size $d=0.6~[\sqrt{\hbar/m\omega_0}]$ and the grid number $n=30$ are chosen along the $x$, $y$ and $z$ directions.
The distribution of the grids in each direction is symmetric with respect to the origin of the coordinate system,
and the spatial derivatives are performed in the momentum space with the fast Fourier transformation~\cite{maruhn2014tdhf,ren2017solving}.


\section{Problems in the 3D lattice calculation}\label{sec3Dchall}

One has to deal with several numerical problems when solving the Nilsson Hamiltonian~\eqref{H} in 3D lattice space by ITM.
They are the $\bm{l}^2$ divergence problem, the calculation of the $\langle l^2\rangle_N$ term, and the convergence problem of the iteration.


\subsection{The $\textbf{l}^2$ divergence problem}

The $\bm{l}^2$ divergence problem resulting from the $D[\bm{l}^2-\langle \bm{l}^2\rangle_N]$ term in Eq.~\eqref{H} occurs in high-$l$ case.
In the spherical Nilsson Hamiltonian, for instance, the energy contribution of $D[\bm{l}^2-\langle \bm{l}^2\rangle_N]$ term to the single-particle state $|Nlj\Omega\rangle$ can be calculated by
\begin{align}
  & D[\bm{l}^2-\langle\bm{l}^2\rangle_N]|Nlj\Omega\rangle \notag\\
 =& \frac{1}{2}D\left[l^2-(4n_r+1)l-6n_r\right]|Nlj\Omega\rangle, ~N=2n_r+l,
\end{align}
where $n_r$ is the radial node number.
Since $D$ is typically negative, the single-particle energy approaches to negative infinity when $l\rightarrow\infty$.

\begin{figure}[H]
  \centerline{
  \includegraphics[width=0.32\textwidth,angle=0]{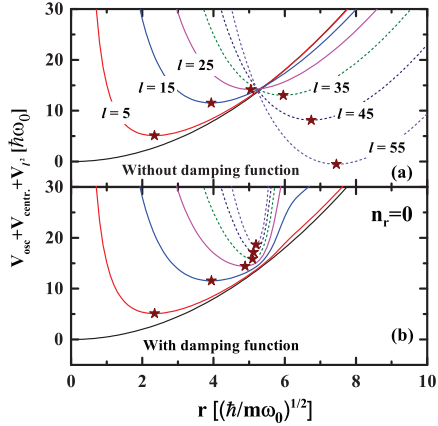}
  } \caption{(Color online)
   (a) Effective radial potential, $V_{\rm eff.} (r)=V_{\rm osc.}(r)+V_{\rm centr.}(r)+V_{\bm{l}^2}(r)$, of the spherical Nilsson Hamiltonian for a given orbital quantum number $l$, when the radial number $n_r=0$;
   (b) Same as panel (a), but the term $V_{\bm{l}^2}(r)$ therein has been modified by a damping function~$f_D(r)$ in Eq.~\eqref{eqfdamp}.
   The minimum of the effective radial potential for a given orbital quantum number is labeled as pentagram.
  }
  \label{fig1}
\end{figure}

To have a closer look, we define the effective radial potential,
\begin{align}
  V_{\rm eff.}(r)
 &=V_{\rm osc.}(r)+V_{\rm centr.}(r)+V_{\bm{l}^2}(r)\notag\\
& =\frac{1}{2}r^2 + \frac{l(l+1)}{2r^2} + \frac{1}{2}D\left[l^2-(4n_r+1)l-6n_r\right],
\end{align}
where the centrifugal potential $V_{\rm centr.}(r)$ and the $D[\bm{l^2}-\langle \bm{l}^2\rangle_N]$ term $V_{\bm{l}^2}(r)$ are included.
Taking radial node number $n_r=0$ as an example, we show $V_{\rm eff.}(r)$ for various values of $l$ in Fig.~\ref{fig1}(a).
From Fig.~\ref{fig1}(a), the minimum of $V_{\rm eff.}(r)$ goes upward with the increase of $l$ when $l<25$. However,
in the high-$l$ cases $(l>25)$, the minimum goes downward and even becomes negative.
Obviously, these high-$l$ states are unphysical solutions, and should not be considered.
In the H.O. basis calculation, these high-$l$ states are excluded automatically due to the truncation of the model space.
In 3D lattice space, however, it takes into account equivalently a large H.O. basis space,
so that these high-$l$ states are included and lead to the divergence problem of the 3D lattice calculation.

From Fig.~\ref{fig1}(a), it can be found that the radial position of the minimum of $V_{\rm eff.}(r)$ increases monotonically with $l$.
Therefore, we propose to restrict the action region of $D\bm{l}^2$ to a moderate space.
In order to realise the restriction of the action region of $D\bm{l}^2$, a Fermi-type damping function is introduced,
\begin{align}
  f_{D}(r) = \frac{1}{1+e^{(r-r_D)/a_D}},\label{eqfdamp}
\end{align}
where the parameter $r_{D}$ is the effective cut-off and $a_{D}$ is the smooth parameter.
Then, $D\bm{l}^2$ is divided into two parts, i.e., $Df_D(r)\bm{l}^2f_{D}(r)$ and $D[\bm{l}^2 - f_{D}(r)\bm{l}^2f_{D}(r)]$,
and the second part is regarded as a perturbation in the calculations.

There are two issues for choosing the damping parameters $r_D$ and $a_D$.
On the one hand, the $V_{\rm eff.}(r)$ of unphysical high-$l$ states should be raised properly.
On the other hand, the effects for the $V_{\rm eff.}(r)$ of physical low-$l$ states should be small.
In the present work, $r_D=6$ and $a_D=0.2$ are adopted. The corresponding effective radial potentials are shown in Fig.~\ref{fig1}(b),
where one can see that the two criterions mentioned above are fulfilled.
The effects of the damping function for the single-particle states $|\psi\rangle$ could be evaluated quantitatively by the first-order correction $\Delta E_{\rm damping}^{(1)} = \langle \psi |\bm{l}^2 - f_{D}(r)\bm{l}^2f_{D}(r)  |\psi\rangle$,
and the detailed discussion in the spherical case can be found in Sec.~\ref{subsec3Daccu}.


\subsection{The calculation of the $\langle \textbf{l}^2\rangle_N$ term }\label{sec_treat_l2n}

The calculation of the $\langle \bm{l}^2\rangle_N$ term is trivial in spherical harmonic basis representation as $N$ is a good quantum number.
However, the calculation of the $\langle \bm{l}^2\rangle_N$ term in 3D lattice space is troublesome since $N$ is not explicitly shown.
To calculate this term in 3D lattice space, we rewrite $\langle \bm{l}^2\rangle_N$ as a $N$-independent operator form.

In the representation of the spherical H.O. basis, $\langle \bm{l}^2\rangle_N$ and $\hat{h}_0=-(\Delta+r^2)/2$ contribute only to the diagonal element with $N(N+3)/2$ and $N+3/2$,
respectively, so $\langle \bm{l}^2\rangle_N$ can be replaced as,
\begin{align}
   \langle \bm{l}^2 \rangle_N
& =\frac{N(N+3)}{2}\notag\\
& =\langle Nlj\Omega|\left(\hat{h}_0-3/2\right) \left(\hat{h}_0+3/2\right)|Nlj\Omega\rangle.
\end{align}
Due to the orthogonal completeness of the basis, $\langle \bm{l}^2 \rangle_N \equiv \hat{h}_0^2-9/2$,
and by replacing the $\langle \bm{l}^2\rangle_N$ term in Hamiltonian with $ \hat{h}_0^2-9/2$ one can calculate it in the 3D lattice space.


\subsection{The improvement of the convergence in the 3D lattice calculation}

Another practical problem is that the potential $V(r,\theta)$ used in Nilsson Hamiltonian~\eqref{H} increases as $r^2$ with $r\rightarrow\infty$,
which appreciably contributes to the states $\{\phi_k\}$ in Eq.~\eqref{phiiter} with high energy,
and makes the iteration in Eq.~\eqref{ekdamp} converging slowly and even running into pathological result.
In Sec.~\ref{sec_treat_l2n}, the potential $V_0(r)=r^2/2$ introduced in $\hat{h}_0^2-9/2$ also leads to the
similar problem.
In order to speed up the convergence of the iteration,
the potential $V(r,\theta)$ and $V_0(r)$ can be modified as $V(r,\theta)=\min\{V(r,\theta),V_{\rm cut1}\}$ and $V_0(r)=\min\{V_0(r),V_{\rm cut2}\}$, respectively.
In the practical calculations, we find that $V_{\rm cut1}=40~[\hbar\omega_0]$ and $V_{\rm cut2}=16~[\hbar\omega_0]$ give a good convergence.
For the energies under $7~[\hbar\omega_0]$, the values of $V_{\rm cut1}=40~[\hbar\omega_0]$ and $V_{\rm cut2}=16~[\hbar\omega_0]$ are reasonable for the description of these levels.


\subsection{Numerical accuracy of the 3D lattice calculation}\label{subsec3Daccu}

In order to examine the efficiency of the numerical techniques introduced above, 3D lattice calculation is used to solve the spherical Nilsson Hamiltonian.
With taking first-order perturbation correction of the damping function, the accuracy of the calculation is checked by comparing the obtained single-particle energies with the exact solutions.

\begin{figure}[H]
  \centerline{
  \includegraphics[width=0.32\textwidth,angle=0]{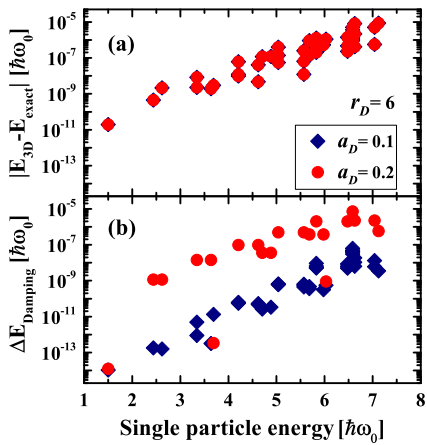}
  } \caption{(Color online)
  For a spherical Nilsson Hamiltonian, the absolute deviations of the single-particle energies obtained by the 3D lattice calculation from the exact solutions
  (a) and the first-order corrections to the single-particle energies introduced by the damping function
  (b) as functions of the single-particle energy.
  The red dots and the blue lozenges represent the results with the damping factors $a_D$=0.1 and 0.2, respectively.
  }
  \label{fig2}
\end{figure}

As shown in Fig.~\ref{fig2}(a), the absolute deviations of the single-particle energies between the 3D lattice calculation and the exact solutions are given as functions of the single-particle energy for $a_D=0.1$ and 0.2 cases, respectively.
Meanwhile, the corresponding first order corrections $\Delta E_{\rm damping}^{(1)}$ are shown in Fig.~\ref{fig2}(b).

It can be seen that the absolute deviations between the 3D lattice calculation and the exact solutions are smaller than $10^{-5}[\hbar\omega_0]$ for $a_D=0.1$ and 0.2 cases, respectively.
This indicates that the high accuracy of the numerical techniques introduced in Sec.~\ref{sec3Dchall}.
From Fig.~\ref{fig2}(b), the first-order corrections $\Delta E_{\rm damping}^{(1)}$ are smaller than $10^{-5}[\hbar\omega_0]$ and $10^{-7}[\hbar\omega_0]$ for $a_D=0.1$ and 0.2 cases, respectively,
and they are also much less than the single-particle energies.
It is reasonable to regard $D[\bm{l}^2 - f_{D}(r)\bm{l}^2f_{D}(r)]$ as a perturbation term.
Comparing Fig.~\ref{fig2}(a) and \ref{fig2}(b), one can find that the decreases of
$\Delta E_{\rm damping}^{(1)}$ are change by two orders of magnitude from $a_D=0.2$ to 0.1,
while the accuracy of the 3D lattice calculation are nearly
unchanged. Therefore, one can conclude that the deviations in Fig.~\ref{fig2}(a) are not from the introduction of the damping function.


\section{Single particles in a reflection-asymmetric potential}\label{sec_sp}

In this section, the single particles in a reflection-asymmetric potential are solved by the ITM in 3D lattice space with the numerical techniques mentioned in Sec.~\ref{sec3Dchall}.
As comparisons, the corresponding Schr\"{o}dinger equations of the Hamiltonian are also diagonalized by using the spherical Harmonic oscillator basis.

\begin{figure}[H]
  \centerline{
  \centering
  \includegraphics[width=0.42\textwidth,angle=0]{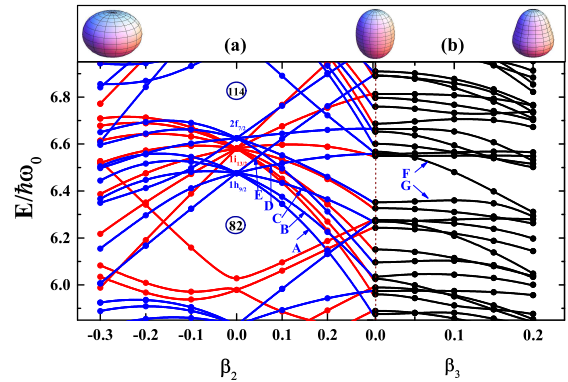}
  } \caption{(Color online)
  The single-particle levels, obtained by the 3D lattice calculation (dots) and the H.O. basis calculation (lines),
  as functions of the deformation parameters $\beta_2$ and $\beta_3$.
  The red (blue) represents positive (negative) parity.
  The shapes shown in the top panel correspond to the deformation parameters $(\beta_2,\beta_3)=(-0.3,0),(0.3,0),(0.3,0.2)$, respectively.
  }\label{fig3}
\end{figure}

In Fig.~\ref{fig3}, we show the single-particle levels as functions of the quadrupole deformation $\beta_2$ on the left panel and the octupole deformation $\beta_3$ with $\beta_2$ fixed on the right panel.
The lines with various colors represent the results obtained by H.O. basis calculation,
and the red (blue) lines denote the levels with positive (negative) parity.
The dots represent the results obtained by 3D lattice calculation.
It can be seen that the agreement between 3D lattice calculation and H.O. basis calculation is very satisfactory.
For the single-particle levels shown in Fig.~\ref{fig3}, all the differences between these two numerical methods are smaller than $10^{-4}~[\hbar\omega_0]$.
This demonstrates the high accuracy and good reliability of these two numerical methods.

For the reflection-symmetric case ($\beta_3=0$) as shown in Fig.~\ref{fig3}(a),
it can be found that the calculations give a sizeable energy gap at proton numbers 82 and 114 in the spherical case.
This reproduces the traditional proton magic number 82 and suggests the proton magic number 114 for superheavy
nuclei.
There are three spherical orbits, i.e., $1h_{9/2}$, $1i_{13/2}$, $2f_{7/2}$, between these two gaps. With the quadrupole deformation $\beta_2$,
the degeneracy due to the spherical symmetry is removed, and the energy gaps at 82 and 114 are quenched.
In order to investigate the evolution of the deformed single-particle levels, the levels coming from $1h_{9/2}$ are chosen,
where the energy Levels A$\sim$E denote the levels with $z-$components of total angular moments $\Omega=1/2\sim9/2$.
It can be found that, for $\beta_2>0$ (prolate shape), the energy Levels A$\sim$E drop rapidly with $\beta_2$
for low $\Omega$ ones,
and increase sharply with $\Omega$. For $\beta_2<0$ (oblate shape), the opposite feature of these levels are shown in the small deformation region,
while all levels show a decrease trend in a larger deformation region.
The slopes of the energy Levels A$\sim$E could be evaluated by the first-order perturbation correction of the quadrupole operator $q = r^2Y_{20}$~\cite{ring2004nuclear}.

\begin{table*}[htbp]
  \caption{Single-particle energy Levels F and G together with the corresponding contributions from the five leading components for $(\beta_2,\beta_3)=(0.3,0),(0.3,0.1),(0.3,0.2)$, respectively.}
  \scalebox{0.8}{
  \setlength{\tabcolsep}{8pt}
  \begin{tabular}{ccllllll}
     \hline
     \hline
     ($\beta_2$,$\beta_3$)  &  Level &~~~~1st comp. &~~~~2nd comp. &~~~~3rd comp. &~~~~4th comp. &~~~~5th comp.\\
     \hline
 (0.3,0.0) &  F     & $[5,3,\frac{5}{2},\frac{1}{2}]$ ~ $30.7\%$
                    & $[5,1,\frac{1}{2},\frac{1}{2}]$ ~ $20.9\%$
                    & $[5,3,\frac{7}{2},\frac{1}{2}]$ ~ $17.2\%$
                    & $[5,5,\frac{9}{2},\frac{1}{2}]$ ~ $13.6\%$
                    & $[7,7,\frac{15}{2},\frac{1}{2}]$~ $6.4\%$\\
           &  G     & $[5,3,\frac{7}{2},\frac{3}{2}]$~~  $65.2\%$
                    & $[5,5,\frac{9}{2},\frac{3}{2}]$~~  $9.4\%$
                    & $[5,1,\frac{3}{2},\frac{3}{2}]$~~  $9.4\%$
                    & $[5,5,\frac{11}{2},\frac{3}{2}]$~ $6.1\%$
                    & $[7,5,\frac{11}{2},\frac{3}{2}]$~  $2.7\%$\\
    \hline
 (0.3,0.1) &  F     & $[7,7,\frac{15}{2},\frac{1}{2}]$~ $18.0\%$
                    & ${\bm [5,3,\frac{7}{2},\frac{1}{2}]}$~ ${\bm 13.4\%}$
                    & $[5,5,\frac{9}{2},\frac{1}{2}]$~~ $11.3\%$
                    & $[5,3,\frac{5}{2},\frac{1}{2}]$~~ $8.8\%$
                    & ${\bm [6,6,\frac{13}{2},\frac{1}{2}]}$ ${\bm 8.4\%}$\\
           &  G     & $ [5,3,\frac{7}{2},\frac{3}{2}]$~~ $60.1\%$
                    & $[5,1,\frac{3}{2},\frac{3}{2}]$~~ $9.0\%$
                    & $[5,5,\frac{9}{2},\frac{3}{2}]$~~ $7.7\%$
                    & $[7,7,\frac{15}{2},\frac{3}{2}]$~ $5.1\%$
                    & $[5,5,\frac{11}{2},\frac{3}{2}]$~ $2.9\%$\\
    \hline
(0.3,0.2)  &  F     & ${\bm[6,6,\frac{13}{2},\frac{1}{2}]}$ ${\bm33.2\%}$
                    & $[5,5,\frac{9}{2},\frac{1}{2}]$~~ $13.1\%$
                    & $[7,7,\frac{15}{2},\frac{1}{2}]$~ $9.4\%$
                    & ${\bm[5,3,\frac{7}{2},\frac{1}{2}]}$~ ${\bm6.3\%}$
                    & $[4,2,\frac{5}{2},\frac{1}{2}]$~~ $2.7\%$\\
           &  G     & ${\bm[5,3,\frac{7}{2},\frac{3}{2}]}$~ ${\bm38.9\%}$
                    & $[7,7,\frac{15}{2},\frac{3}{2}]$~ $12.9\%$
                    & ${\bm[6,6,\frac{13}{2},\frac{3}{2}]}$ ${\bm5.9\%}$
                    & $[5,5,\frac{9}{2},\frac{3}{2}]$~~ $5.7\%$
                    & $[5,5,\frac{3}{2},\frac{3}{2}]$~~ $4.8\%$\\
    \hline
    \hline
\end{tabular}}\label{tab_com}
\end{table*}
For the reflection-asymmetric case ($\beta_3\neq0$) as shown in Fig.~\ref{fig3}(b), the evolution of most energy levels with respect to $\beta_3$ is rather slow.
Due to the odd parity of the operator $r^2Y_{30}$, its first order perturbation correction vanishes.
The slopes of the energy levels with respect to $\beta_3$ could be evaluated by the second order perturbation correction,
which give rather gentle slopes compared to the quadruple deformed case.
However, some levels show more drastic change with increase of $\beta_3$.
It can be understood by analysing the octupole coupling between the energy levels.
It is well known that the condition for strong octupole coupling is driven by the interaction between the orbitals $(l,j)$ and $(l\pm3,j\pm3)$ through the $Y_{30}$ potential~\cite{bohr1953collective}.
In order to understand the evolution of the energy levels with the deformation $\beta_3$, the energy Levels F and G coming from $2f_{5/2}$ and $2f_{7/2}$ are chosen.
These levels together with their contributions from the five leading components (labeled by the spherical quantum numbers $[Nlj\Omega]$) for the $(\beta_2,\beta_3)=(0.3,0),(0.3,0.1),(0.3,0.2)$ cases are shown in Table~\ref{tab_com}, respectively.
For the energy Level F with a drastically decreasing tendency, its second ($13.4\%$) and third ($8.4\%$) components compose
an octupole deformation driving pairs $2f_{7/2}[5,3,\frac{7}{2},\frac{1}{2}]$ and $i_{13/2}[6,6,\frac{13}{2},\frac{1}{2}]$ in the $(\beta_2,\beta_3)=(0.3,0.1)$ case.
The weights of this driving pair grow with the increase of $\beta_3$, and a successive decrease in the energy Level F is given.
For the energy Level G with a slowly decreasing tendency, no octupole deformation driving pair is found among the five leading components in the $(\beta_2,\beta_3)=(0.3,0.1)$ case.
With the increase of $\beta_3$, the driving pair $f_{7/2}[5,3,\frac{7}{2},\frac{1}{2}]$ and $i_{13/2}[6,6,\frac{13}{2},\frac{1}{2}]$ composed by the first ($38.9\%$) and second ($5.9\%$) components appears, and the energy Level G shows an obvious decreasing tendency.
Therefore, the tendency of the energy levels with the octupole deformation can be well understood by the octuple coupling between the driving pairs.

\begin{figure}[H]
  \centerline{
  \includegraphics[width=0.42\textwidth,angle=0]{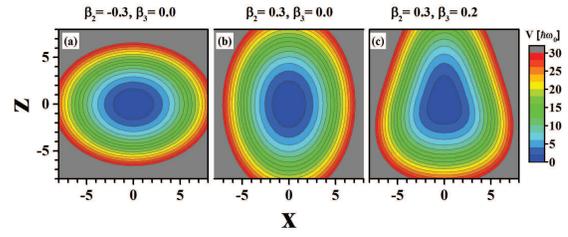}
  } \caption{(Color online)
  Contour plots of the reflection-asymmetric potentials on the plane determined by the $x$-axis and the axially symmetric $z$-axis.
  The deformation parameters corresponding to (a), (b) and (c) are $(\beta_2,\beta_3)=(-0.3,0),(0.3,0),(0.3,0.2)$, respectively.
  }
  \label{fig4}
\end{figure}
\begin{figure*}[htbp]
  \centerline{
  \includegraphics[width=0.74\textwidth,angle=0]{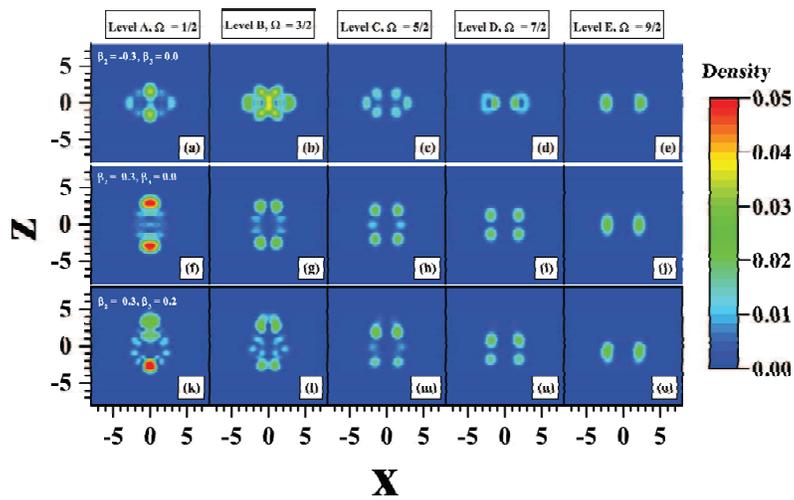}
  } \caption{(Color online)
  The density distributions versus $x$ and $z$ on the $y = 0.3$ plane for the energy Levels A-E shown in Fig.~\ref{fig3} obtained by the 3D lattice calculations.
  The deformation parameters are $(\beta_2,\beta_3)=(-0.3,0),(0.3,0),(0.3,0.2)$, respectively.
  }
  \label{fig5}
\end{figure*}
Furthermore, all these evolution features can be understood by comparing the shapes of the potentials and the density distributions of the single-particle levels.
Fig.~\ref{fig4} displays the distributions of the reflection-asymmetric potentials with $(\beta_2,\beta_3)=(-0.3,0)$, $(0.3,0)$, $(0.3,0.2)$ in the $y=0$ plane.
The corresponding single-particle densities of the energy Levels A$\sim$E in the $xz$ plane at the positions of grid point in the $y$ directions closest to zero,
namely, at the point of $y=d/2=0.3$, are depicted in Fig.\ref{fig5}.

The quadrupole filed $r^2Y_{20}$ causes the levels with lower $\Omega$ values to be shifted downwards for positive deformations (prolate shapes) and to be shifted upwards for negative deformations (oblate shapes).
One can understand this effect, realizing that the states with lower $\Omega$-values have a relatively high probability of being close to the $z$-axis.
Comparing Fig.~\ref{fig4}(a) and \ref{fig4}(b), it can be found that the potential gets softer in the $z$-direction and steeper in the $xy$-directions with the increase of $\beta_2$.
The single-particle density distributions for the energy Levels A$\sim$ E in the $(\beta_2,\beta_3)=(0.3,0)$ case are shown in the second row of Fig.~\ref{fig5}.
The density distribution of the energy Level A is mainly near the $z$-axis, and a softer potential in the $z$-direction moves down the energy of this level.
Similarly, the density distribution of the energy Level E is mainly near the $z=0$ plane, and a steeper potential in the $xy$-directions moves up the energy of this level. According to the above discussions,
the evolutions of the energy Levels A$\sim$ E with respect to $\beta_2$ shown in Fig.~\ref{fig3}(a) could be understood properly.
For the $(\beta_2,\beta_3)=(-0.3,0)$ case, all the single-particle density distributions of the energy Levels A$\sim$E have considerable parts distributed closely to the $z=0$ plane as shown in the first row of Fig.~\ref{fig5}.
Therefore, the energy Levels A$\sim$E go downwards with the decrease of $\beta_2$ in the
large oblate deformed region.

Comparing Fig.~\ref{fig4}(b) and \ref{fig4}(c), the potential changes rather slightly,
and it is nearly unchanged in the $z$-direction and gets steeper (softer) in the $xy$-directions for $z>0~(z<0)$ with the increase of $\beta_3$.
The density distributions of the energy Levels A$\sim$E in the $(\beta_2,\beta_3)=(0.3,0.2)$ case are shown in the third row of Fig.~\ref{fig5}.
Comparing them with the second row of Fig.~\ref{fig5}, one should note that the change of density distributions is not so drastic, which is consistent with the slow evolution with the increase of $\beta_3$.


\section{Summary and Perspectives}\label{sec_summary}

In summary, the imaginary time method in 3D lattice space and the harmonic oscillator basis expansion method have been applied
to investigate the single-particle motion in a reflection-asymmetric potential by solving the reflection-asymmetric Nilsson Hamiltonian.
Several numerical problems occur when the Nilsson Hamiltonian is solved in 3D lattice space, and have been discussed and overcome with the following numerical techniques:
(i)~the $\bm{l}^2$ divergence problem is solved by introducing a damping function;
(ii)~the difficulty in calculating $\langle \bm{l^2}\rangle_N$ term due to the missing of the good quantum number $N$ is overcome by introducing an equivalent $N$-independent operator.
The efficiency of these numerical techniques is demonstrated by solving the spherical Nilsson Hamiltonian in 3D lattice space.
The consistency of the 3D lattice calculation and the harmonic oscillator basis expansion method is demonstrated by that the differences in the obtained single-particle energy are smaller than 10$^{-4}[\hbar\omega_0]$.
The change of the energy levels with the quadrupole and octupole deformations can be well understood
by the first order perturbation correction of the quadrupole operator $q=r^2Y_{20}$ and the octuple coupling between the octupole deformation driving orbits, respectively.
The evolution of the deformed single-particle levels can be understood from the single-particle density and the reflection-asymmetric potential.

Based on the present work, the reflection-asymmetric particle rotor model is expected to be developed.
So far, the existing reflection-asymmetric particle rotor model is for axially deformed nuclei only~\cite{zaikin1966octupole,leander1984intrinsi}.
With further including the triaxial degree of freedom, the reflection-asymmetric particle rotor model is expected to be developed
and applied to the observed multiple chiral doublet bands with octupole correlations in $^{78}$Br in the near future~\cite{liu2016evidence}.


\Acknowledgements{The authors are indebted to Prof. Jie Meng for constructive guidance and valuable suggestions. Fruitful discussions with Prof. Shuangquan Zhang and Dr. Pengwei Zhao are highly acknowledged.
This work was supported in part by the Major State 973 Program of China (Grant No. 2013CB834400), the National Natural Science Foundation of China (Grants No. 11335002, No. 11375015, No. 11461141002, and No. 11621131001).}

\InterestConflict{The authors declare that they have no conflict of interest.}



\end{multicols}
\end{document}